%% file: acmart-sigplanproc-template.tex
\documentclass[sigplan]{acmart}

\acmConference[arXiv'19]{arXiv Paper}{2019}{USA}
\acmYear{2019}
\acmISBN{} 
\acmDOI{} 
\startPage{1}

\setcopyright{none}

\usepackage{booktabs}   
\usepackage{subcaption} 

\input{00-my-config.tex}

\begin{document}


\title[Microbenchmarking Deep Copy]{Assessing Performance Implications of Deep Copy Operations via Microbenchmarking}         


\author{Millad Ghane}
\affiliation{
  \department{Department of Computer Science}              
  \institution{University of Houston}            
  \state{TX}
  \country{USA}                    
}
\email{mghane2@uh.edu}          

\author{Sunita Chandrasekaran}
\affiliation{
  \department{Department of Computer and Information Sciences}             
  \institution{University of Delaware}           
  \state{DE}
  \country{USA}                   
}
\email{schandra@udel.edu}         

\author{Margaret S. Cheung}
\affiliation{
  \department{Physics Department}             
  \institution{University of Houston}           
}
\affiliation{
  \department{Center for Theoretical Biological Physics, Rice University}             
  \state{TX}
  \country{USA}                   
}
\email{mscheung@central.uh.edu}         

\begin{abstract}
\input{00-abstract}

\end{abstract}

\ccsdesc[500]{Computer systems organization~Heterogeneous (hybrid) systems}
\ccsdesc[500]{Computer systems organization~High-level language architectures}

\newcommand{\red}[1]{\textcolor{red}{#1}}

\keywords{Deep Copy, Memory Subsystem, Benchmark, Heterogeneous, Portability, Productivity, Programming Model}  

\settopmatter{printfolios=true}

\maketitle

\input{01-intro}

\input{02-deepcopy}

\input{03-pointerchain.tex}

\input{04-model}

\input{05-evaluation}

\input{06-result}

\input{07-relatedwork}
\input{08-conclusion}

\begin{acks}                            
  This material is based upon work supported by the
  \grantsponsor{GS100000001}{National Science
    Foundation}{https://www.nsf.gov/awardsearch/showAward?AWD_ID=1531814} under Grant~\grantnum{GS100000001}{OAC-1531814} and \grantnum{GS100000001}{MCB-1412532}, and \grantsponsor{GS100000002}{Department of Energy}{} under Grant~\grantnum{GS100000002}{DE-SC0016501}. 
\end{acks}

\bibliography{acmart-sigplanproc-template.bib}



\end{document}

%% file: 00-my-config.tex
\usepackage{tikz}

\usepackage{algorithm,algpseudocode}

\algrenewcommand\algorithmicrequire{\textbf{Input:}}
\algrenewcommand\algorithmicensure{\textbf{Output:}}
\renewcommand{\algorithmicrequire}{\textbf{Input:}}
\renewcommand{\algorithmicensure}{\textbf{Output:}}

\algnewcommand\algorithmicforeach{\textbf{for each}}
\algdef{S}[FOR]{ForEach}[1]{\algorithmicforeach\ #1\ \algorithmicdo}

\algnewcommand{\algorithmicand}{\textbf{ and }}
\algnewcommand{\algorithmicor}{\textbf{ or }}
\algnewcommand{\OR}{\algorithmicor}
\algnewcommand{\AND}{\algorithmicand}
\algnewcommand{\var}{\texttt}

\algnewcommand{\algvar}{\texttt}
\algnewcommand{\assign}{\leftarrow}
\algnewcommand{\makeconst}[1]{\textsc{#1}}

\newcommand{\textct}[1]{\noindent \texttt{\textbf{\#\detokenize{#1}}}}

\usepackage{caption}
\usepackage{subcaption}

\newcommand*\justify{%
  \fontdimen2\font=0.4em
  \fontdimen3\font=0.2em
  \fontdimen4\font=0.1em
  \fontdimen7\font=0.1em
  \hyphenchar\font=`\-
}

\usepackage{multirow}

\usepackage{listings}

\usepackage{xcolor}
\usepackage{listings}

\lstdefinestyle{customc}{
  float=t,
  aboveskip=0.5em,
  belowskip=0.5em,
  abovecaptionskip=0.5\baselineskip,
  belowcaptionskip=0.5\baselineskip,
  breaklines=true,
  frame=tb,
  floatplacement=t,
  language=C,
  showstringspaces=false,
  basicstyle=\footnotesize\ttfamily,
  stepnumber=1,
  showstringspaces=false,
  tabsize=2,
  breaklines=true,
  breakatwhitespace=false,
  numbers=left,
  xleftmargin=2em,
  framexleftmargin=2.2em
}

%% file: 00-abstract.tex
As scientific frameworks become sophisticated, so do their data structures. Current data structures are no longer simple in design and they have been progressively complicated. The typical trend in designing data structures in scientific applications are basically nested data structures: pointing to a data structure within another one. Managing nested data structures on a modern heterogeneous system requires tremendous effort due to the separate memory space design. 

In this paper, we will discuss the implications of deep copy on data transfers on current heterogeneous. Then, we will discuss the two options that are currently available to perform the memory copy operations on complex structures and will introduce \textbf{\texttt{pointerchain}} directive that we proposed. Afterwards, we will introduce a set of extensive benchmarks to compare the available approaches. Our goal is to make our proposed benchmarks a base to examine the efficiency of upcoming approaches that address the challenge of the deep copy operation.

%% file: 01-intro.tex
\section{Introduction}
\label{intro}

\begin{figure}[t]
\centering
\includegraphics[width=0.8\columnwidth]{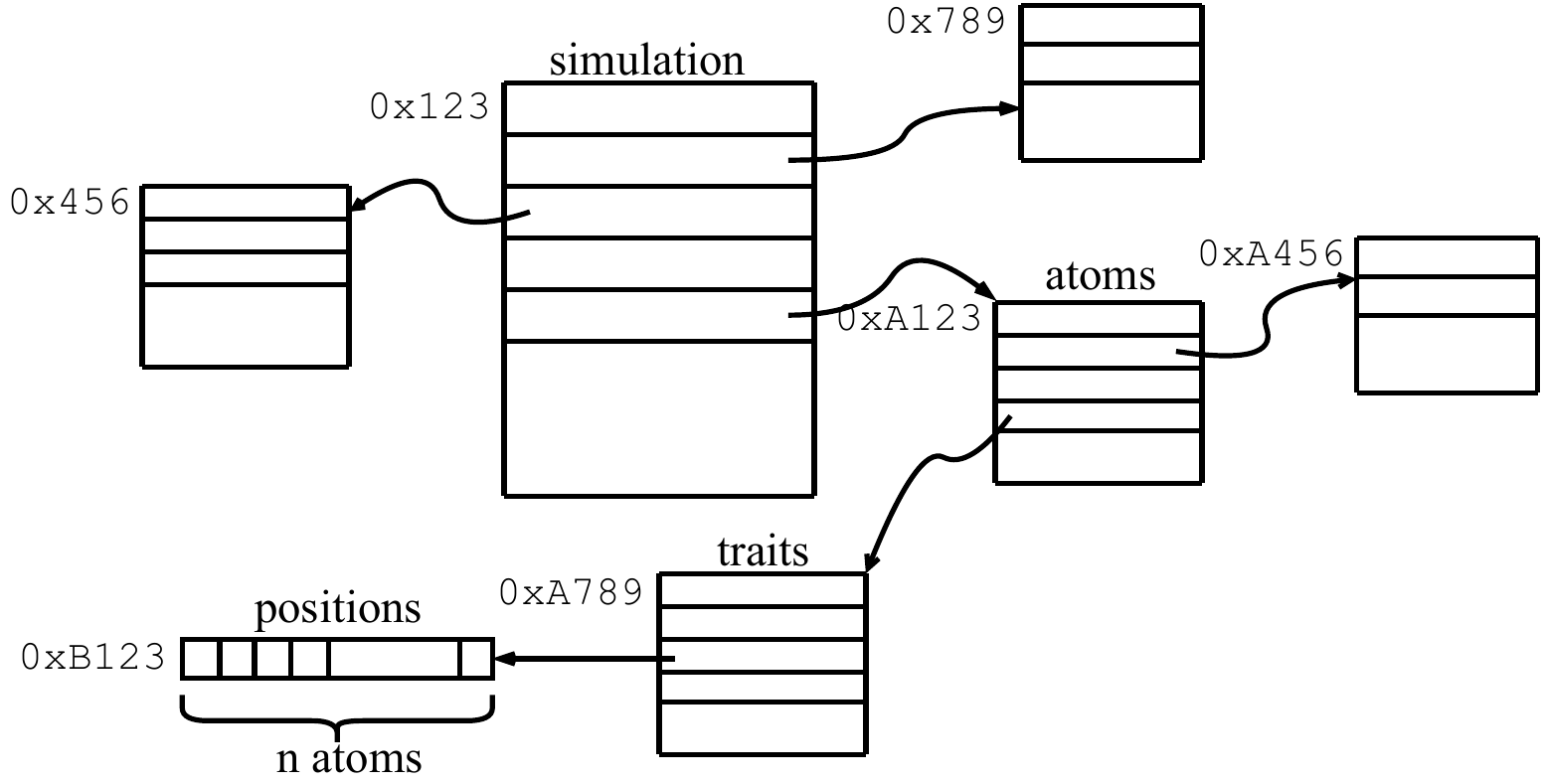}
\caption{An example of a pointer chain. An illustration of a data structure and its children. In order to reach the \texttt{position} array, one must go through a chain of pointers to extract the effective address.}
\label{fig:pointerchain}
\end{figure}

Energy efficiency has been at the forefront of the high-performa\-nce computing~(HPC) developments to tackle energy and power consumption crisis of HPC systems~\cite{subramaniam2013trends}. 
A promising approach that fulfills the DARPA's requirements~\cite{subramaniam2013trends} in designing next generation of exascale supercomputers has been heterogeneity~\cite{keckler2011gpus,lucas2014doe,vetter2017advanced,mittal2015heterosurvey,liu2010survey,ghane2018roofline}; from the node-level heterogeneity like Titan~\cite{titan} and Summit~\cite{ornl2018summit} to the chip-level heterogeneity as in the System-on-a-chip architectures~\cite{ghane2014nocflow,benini2002noc}. 
However, developing applications for heterogeneous systems is not an easy task and requires novel approaches~(e.g., directive-based programming models~\cite{openacc,openmp,ghane2015ompt,ghane2018chameleon,ghane2019gecko}) to assist the application developers in their efforts. 
Heterogeneous systems, on the other hand, have multiple and separate levels of memory spaces, which such design requires developers to explicitly issue data transfers from one memory space to another with a set of software APIs. 
For instance, in a system composed of a host processor and an accelerator, the host processor cannot directly access the data on the device and vice versa. For such systems, the data is copied back and forth between the host and the accelerator. This issue becomes particularly severe for scientific applications as their data structure becomes very complicated.

As a scientific framework becomes sophisticated, so does its data structures. A data structure typically includes pointers (or dynamic arrays) that point to the \textit{primitive data types} or to other user-defined data types. As a result, transferring data structures from the host to the other devices mandates not only the transfer of the main data structure but also its nested data structures. This transfer process is also known as the \textit{deep copy}. The tracking of the pointers that represent the main data structure on the host from its counterpart on the device further complicates the maintenance of the data structure. Although this complicated process of performing the deep copy operation avoids a major change in the source codes, it imposes unnecessary data transfers on the application. In some cases, a \textit{selective deep copy} is sufficient when only a subset of the fields of the data structure on the device is of our interest~\cite{deepcopyTR2016}. However, even though the data motion decreases proportionally, the burden to maintain data consistency among the host and other devices still exists.

Our contributions in this paper are as following:
\begin{itemize}
\item We will discuss the challenges of transferring a nested data structure to the device and the available options to perform the transferring. We also discuss the semantics of deep copy, the required steps to take, and the available options to perform a deep copy operation (Section~\ref{deepcopy}).
\item We will introduce the \textbf{\texttt{pointerchain}} directive (Section~\ref{pointerchain_directive_sec}) as an alternative approach to tackle the challenges imposed by the pointer management of a nested data structure. Our directive reduces the amount of codes generated for the host and the device. 
\item We will design a set of benchmark applications to examine approaches that perform deep copy (Section~\ref{model}). Our design includes two scenarios that benefit from performing deep copy; \textit{Linear} and \textit{Dense} scenarios. In the \textit{Linear} scenario, our targeted array is placed in depth within the nested hierarchy. 
In the \textit{Dense} scenario, the intermediate pointers (e.g., \texttt{atoms} in Figure~\ref{fig:pointerchain}) are an array themselves. This will put a lot of stress on the approach that is going to be examined by our benchmark. 
\item We will discuss the results of our proposed scenarios (Section~\ref{results}) on three separate approaches to deep copy: Unified Virtual Memory~(UVM)~\cite{landaverde2014uvm}, marshalling/unmarshalling the data structure tree, and \textbf{\texttt{pointerchain}}. 
\end{itemize}

%% file: 02-deepcopy.tex
\section{Semantics of Deep Copy}
\label{deepcopy}


Memory spaces in modern HPC platforms are categorized into two separate spaces: the \textit{host} memory space and the \textit{device} memory space. A memory allocation in one space does not guarantee an allocation in the other. In order to guarantee data consistency, such an approach requires a complete replication of all data structure in both spaces. However, data structures get complicated as they preserve the complex states of an application.

Figure~\ref{fig:pointerchain} illustrates a common case in the design of a data structure for scientific applications. The arrows represent pointers. The number next to each structure shows the potential physical address of an object in the memory. The main data structure is the \texttt{simulation} structure. Each object of this structure has pointers embedded to the other structures, in this case, the \texttt{atoms} structure. The \texttt{atoms} structure also has a pointer to another \texttt{traits} structure, and so on. Therefore, in order to access the elements of the \texttt{positions} array starting from the \texttt{simulation} object, we have to dereference the following chain of pointers: \texttt{\justify simulation->atoms->traits->positions}. Every arrow from this chain goes through a dereference process to extract the effective address of the final pointer. We call this chain of accesses to reach the final pointer (in this case, \texttt{positions}) a \textit{pointer chain}. Since every pointer chain eventually resolves to a memory address, we proposed the extraction of the effective address and replace it with their corresponding pointer chain in the parallel sections of the code.

There are two primary techniques to efficiently utilize pointer chains within the source code.
The first technique is the deep copy that requires excessive data transfer between the host and the device. The second technique is the utilization of Unified Virtual Memory~(UVM) on Nvidia devices~\cite{landaverde2014uvm}. UVM provides a single coherent memory image to all processors (CPUs and GPUs) in the system, which is accessible through a common address space. UVM eliminates the necessity of explicit data transfers by applications. Although it is an effortless approach for developers, it has several drawbacks: 1)~It is only supported by Nvidia devices; 2)~It is not a \textit{performance-friendly} approach due to its arbitrary memory transfers. 
The consistency protocol in UVM depends on the underlying hardware and device driver that traces memory page-faults on both host and device memories. Whenever a page fault occurs on the device, the CUDA~\cite{cuda,farber2011cuda} driver fetches the most up-to-date version of the page from the main memory and provides it to the GPU. A page fault on the host follows similar steps to fetch the updated page from the device.



\begin{figure}[t]
\centering
\includegraphics[width=1.0\columnwidth]{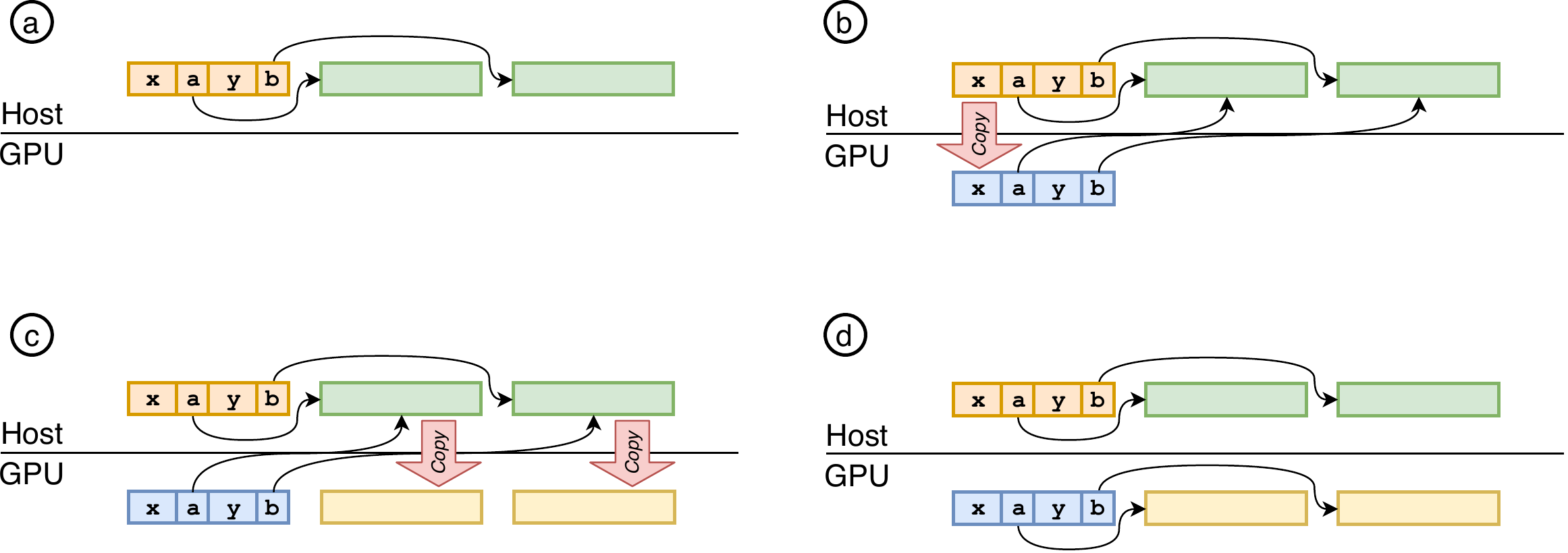}
\caption{Steps to perform a deep copy operation when the targeting device is a GPU. The horizontal line separates the memory spaces between the host and the GPU. (a) initialize the data structures; (b) copy the main structure to the GPU; (c) copy other nested data structures to the device; (d) fix corresponding pointers in every data structure.}
\label{deep_copy_steps}
\vspace{-1em}
\end{figure}

As discussed above, the scientific applications utilize nested data structures in their design. Any data structure (in C/C++) is composed of a set of simple or complex member variables. The simple member variables are those members with primitive data types (e.g., \texttt{int}, \texttt{float}, \texttt{double} in C/C++). However, the complex member variables are those that are user-defined data structures themselves. The situation gets complicated when the complex member variable itself possesses another complex data structure. The common approach to utilize complex member variables in C/C++ for such cases is to define them as pointers. Since the array size is not known at the compilation time, they have to be allocated at the run time. This makes their address in memory to be known only then. This is not an ideal case for heterogeneous platforms with separate memory address spaces. Figure~\ref{deep_copy_steps} illustrates the necessary steps required to perform the deep copy. After initializing (Step~\textit{\textbf{a}}) and transferring (Step~\textit{\textbf{b}} and \textit{\textbf{c}}) the structure from the host to the device, the pointers on the main structure hold illegal addresses. They still point to the same memory address on the host, which is inaccessible on the device. We have to fix this issue by reassigning the pointers to their correct corresponding addresses on the device (Step~\textit{\textbf{d}} in Figure~\ref{deep_copy_steps}).

Deep copy, as described in~\cite{deepcopyTR2016}, can be categorized into two groups: 1)~Full Deep Copy; 2)~Selective (Partial) Deep Copy. A full deep copy operation copies a data structure with all of its nested data structure to the device. As a result, a replica of the whole structure is available on the device. The process discussed in Figure~\ref{deep_copy_steps} demonstrates a full deep copy operation. However, a full deep copy is not always an appropriate approach and we need mechanisms to perform a partial copy operation. In those cases, not all variable members of a data structure are accessed during a kernel execution on the device. As a result, there is no need to transfer them to the device. Consider the example in Figure~\ref{deep_copy_steps}. If our kernel is only accessing array~\textbf{\texttt{x->a}}, we should not copy array~\textbf{\texttt{x->b}} to the device and keep it on the host. This will significantly improve performance of the copy operation. This is an example of a selective deep copy operation. 

Our proposed approach, which we call \textbf{\texttt{pointerchain}}~\cite{ghane2019pointerchain}, is a directive-based approach that provides selective accesses to data fields of a nested data structure while minimizing error opportunities and changes to the source code. 
A brief description of \textbf{\texttt{pointerchain}} is provided in Section~\ref{pointerchain_directive_sec}. For a detailed discussion, please refer to~\cite{ghane2019pointerchain}.

%% file: 03-pointerchain.tex
\section{The \texttt{pointerchain} directive }
\label{pointerchain_directive_sec}

A chain of pointers, similar to the example shown in Figure~\ref{fig:pointerchain}, will be extracted to a set of machine instructions to correctly extract the effective address of the chain for both the host and the device. 
However, dereferencing each intermediate pointer in the chain is the equivalent of two memory load instructions, which are high cost operations. As the pointer chain lengthens with a growing number of intermediate pointers, the program have to perform excessive memory load operations to extract the final effective address that points to the final member of the chain. This extraction process impedes performance, especially when this process (dereferencing a chain of pointers) is happening within a loop (e.g., a \texttt{for}-loop). In order to alleviate the implications of the extraction process, we propose to perform the extraction process before the computation region begins, and then reuse the extracted address within the region afterwards.

We demonstrate the idea behind the extracting process from a pointer chain using the example in Figure~\ref{fig:pointerchain}. 
In this setup, we replace the pointer chain of \texttt{\justify simulation->atoms->traits->positions} with its corresponding effective address, in this case, the memory address of \texttt{positions} array (\texttt{0xB123}) as shown in Figure~\ref{fig:pointerchain}. 
We utilize this address for future data transfers to and from the device and also the computational regions. It prevents transferring redundant data structures (in this case, \texttt{simulation}, \texttt{atoms}, and \texttt{traits}) to the device, which, in any case, will remain intact on the device. The code executed on the device will modify none of these objects. Moreover, it keeps the device busy performing ``useful'' work rather than spending time on extracting effective addresses from the chain.

The effective address utilization, as a replacement to a pointer chain, however, demands code modifications on both the data transfer clauses and the kernel codes. To address these concerns, we propose a set of directives that leads to minimal code changes. 

\input{include/listing-2.tex}

\subsection{Expanded Version}

In its simple form, the \textbf{\texttt{pointerchain}} directive accepts two constructs: \textbf{\texttt{declare}} and \textbf{\texttt{region}}. Developers use \textbf{\texttt{declare}} construct to announce the pointer chains in their code. The syntax in C/C++ is as following:

\vspace{8pt}
\textct{pragma pointerchain declare(}\textit{variable [,variable]...}\texttt{\textbf{)}} 
\vspace{8pt}

\noindent 
where \textit{variable} is defined as below:

\vspace{8pt}
\textbf{\texttt{variable := name\big\{type[:\textit{qualifier}]\big\} }}
\vspace{8pt}

\noindent
where

\begin{itemize}
\item \textbf{\texttt{name}}: the pointer chain 
\item \textbf{\texttt{type}}: the data type of the effective address
\item \textbf{\texttt{qualifier}}: an optional parameter that is either \textbf{\texttt{\justify restrictconst}} or \textbf{\texttt{restrict}}. They will make the underlying variable to be decorated with \textbf{\texttt{\_\_restrict const}} and \textbf{\texttt{\_\_restrict}} in C/C++, respectively. These qualifiers provide hints to the compiler to optimize the code with regard to the effects of pointer aliasing.
\end{itemize}

After declaring the pointer chains in our code, we have to determine the code region that we target to perform the transformation. 
The following lines describe how to use \textbf{\texttt{begin}} and \textbf{\texttt{end}} clauses with \textbf{\texttt{region}} construct. 
The pointer chains that have been declared before in the current scope are the subject for transformation in subsequent regions.

\vspace{8pt}

\textct{pragma pointerchain region begin}

\textit{\texttt{<...computation or data movement...>}}

\textct{pragma pointerchain region end} 

\vspace{8pt}

Our proposed directive, \textbf{\texttt{pointerchain}}, is a language- and programming-model-agnostic directive.
Although, in this paper, for implementation purposes, \textbf{\texttt{pointerchain}} is targeting C/C++ and OpenACC~\cite{openacc} programming models, one can utilize it for the Fortran language or target the OpenMP~\cite{openmp} programming model as well.

\subsection{Condensed Version}

Our two proposed clauses~(\textbf{\texttt{declare}} and \textbf{\texttt{region}}) provide developers with the flexibility of reusing multiple variables in multiple regions. However, there exists a condensed version of \textbf{\texttt{pointerchain}} that performs the declaration and replacement process at the same time. 
The condensed version of \textbf{\texttt{pointerchain}} replaces the declared pointer chain with its effective address in the scope of the targeted region. It is placed on the \textbf{\texttt{region}} clauses. An example of a simplified version, enclosing a computation or data movement region, is shown below:

\vspace{8pt}

\textct{pragma pointerchain region begin declare(}\textit{variable [,variable]...}\texttt{\textbf{)}} 

\textit{\texttt{<...computation or data movement...>}}

\textct{pragma pointerchain region end} 

\vspace{8pt}

The condensed version is a favorable choice in comparison to the \textbf{\texttt{declare/region}} pair when our kernels (regions) have a few variables and we do not reuse the chains in future. It leads to a clean, high quality code. Furthermore, utilizing the pair combination helps with the code readability, reduces the complexity of code, and expedites the porting process to OpenACC and OpenMP programming models. Potentially, the current modern compilers will be able to incorporate the condensed version of \textbf{\texttt{pointerchain}} with the OpenACC or OpenMP directives directly. The following example shows how the condensed version could be incorporated into the OpenACC programming model.

\vspace{8pt}

\textct{pragma acc parallel} \textit{\textbf{pointerchain}(\textit{variable [,variable]...})}

\textit{\texttt{<...computations...>}}

\vspace{8pt}

\subsection{Sample Code}

Listing~\ref{pointerchainexample2} shows an example on how to use \textbf{\texttt{pointerchain}} in a source code. 
Lines 1-16 show the data structures for configuration in Figure~\ref{fig:pointerchain}, including the main object variable (\texttt{simulation}). Our computational kernel, Lines 25-32, initializes the position of every atom in 3D space in the system. These lines represent a normal, formal~\texttt{for-loop} that has been parallelized by the OpenACC programming model. First, we declared our pointer chain (Line~18), then utilized the \textbf{\texttt{region}} clause to transfer the data to our target device (Lines~20-22), and finally, utilized the \textbf{\texttt{region}} clause to parallelize the \texttt{for} loop (Lines~25-32). Without \textbf{\texttt{pointerchain}}, parallelizing the \texttt{for}-loop requires to transfer every member of the chain to the device separately while retaining their relationship during the transfer. This will adversely impact the performance while making its implementation also challenging.


\textbf{\texttt{Pointerchain}} is capable of dealing with both pointers and scalar variables. 
Unlike pointers, dealing with the scalar variables requires more attention. Following example lays out the challenges we encounter in dealing with scalar variables.
Suppose we want to change the number of atoms in the \texttt{atoms} structure (\texttt{\justify simulation->atoms->N}). The \texttt{declare} clause extracts the value stored in this variable and records it in a temporary variable for the future references in the upcoming regions. However, when the region is done, the temporary variable has the most up-to-date value and while its corresponding chain is unaware of such update. Therefore, \textbf{\texttt{pointerchain}} updates the corresponding pointer chain with the updated temporary variable. 


\subsection{Implementation Strategy}

We have developed a Python script that performs a \textit{source-to-source} transformation of the source codes annotated with the \textbf{\texttt{pointerchain}} directives. 
Our transformation script searches for all source files in the current folder, finds those annotated with the \textbf{\texttt{pointerchain}} directives, and then, transforms each \textbf{\texttt{pointerchain}} directive to its equivalent code.

Here is an overview of the transformation process. Upon encountering a \textbf{\texttt{declare}} clause, for each variable, a \textit{local variable} is declared and initialized to the effective address of our corresponding pointer chain. If any \textbf{\texttt{qualifier}} is set for a chain, they will also be appended to the declaration. Any occurrences of the pointer chains in between \textbf{\texttt{region begin}} and \textbf{\texttt{region end}} clauses are replaced with their counterpart local pointers that were declareed in the same functional unit.



%% file: include/listing-2.tex
\begin{lstlisting}[style=customc, label=pointerchainexample2, caption={An example on how to use \textbf{\texttt{pointerchain}} directive for data transfer and kernel execution.}, float]
typedef struct {
	...
	// position, momenta, and force in 3D space
	double *positions[3];
} Traits;
typedef struct {
	...
	// position, momenta, and force in 3D space
	Traits *traits;
} Atoms;
typedef struct {
	...
	// atom data (positions, momenta, ...)
	Atoms* atoms;
} Simulation;
Simulation *simulation;
// Declaring the targeted pointer chain
#pragma pointerchain declare(simulation->atoms->traits->positions{double*})

#pragma pointerchain region begin
#pragma acc data enter copyin(simulation->atoms->traits->positions[0:N])
#pragma pointerchain region end

// pointerchain region
#pragma pointerchain region begin
#pragma acc parallel loop
for(int i=0;i<nAtoms;i++) {
	simulation->atoms->traits->positions[i][0] = ...;
	simulation->atoms->traits->positions[i][1] = ...;
	simulation->atoms->traits->positions[i][2] = ...;
}
#pragma pointerchain region end

\end{lstlisting}

%% file: 04-model.tex
\section{Methodology}
\label{model}

In this section, we will discuss our methodology on benchmarking the deep copy operations for two different scenarios; \textit{Linear} and \textit{Dense}. Each scenario is tested with various \textit{transfer} and \textit{layout} schemes. In the following, we will discuss the detailed description of each scenario and scheme. All the source codes of our microbenchmark are accessible on Github\footnote{https://github.com/milladgit/deepcopy-benchmark}.

\subsection{Linear Scenario}

\begin{figure}[t]
\centering
\includegraphics[width=1.0\columnwidth]{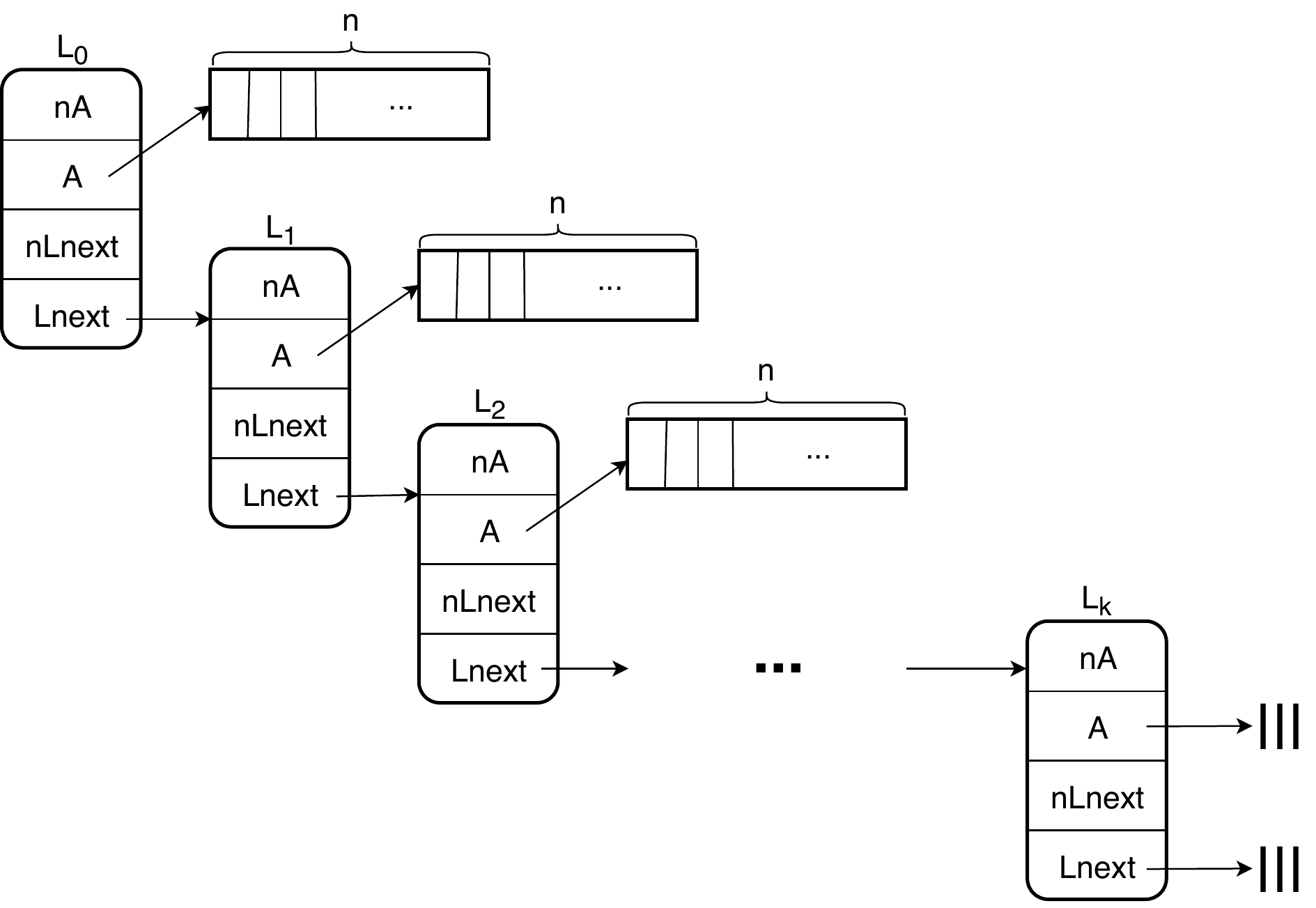}
\caption{The Linear scenario.}
\label{fig:simple_approach_diagram}
\vspace{-1em}
\end{figure}

In the first case, we will design a set of experiments to study the effect of nesting depth on the performance of applications. Figure~\ref{fig:simple_approach_diagram} shows the data layout for the Linear scenario. All the data structures in this scenario have similar member variables. They consist of two integer variables~(\texttt{nA} and \texttt{nLnext}), a floating-point array~(\texttt{A}), and a pointer to the next nested data structure (\texttt{Lnext}). The main data structure is the the data structure at level 0, which is designated with $L_0$. Our design for this scenario has two parameters: \textit{k} and \textit{n}. The parameter \textit{k} controls the depth of our data layout and the parameter \textit{n} controls the length of the extra payload that we have assigned to each nested data structure.

In order to perform these experiments, we developed a Python script that accepts an integer \textit{k} as an input parameter and generates a total of $k$ C++ source files with 1 to \textit{k} nested data structures, similar to the configuration in Figure~\ref{fig:simple_approach_diagram}. The parameter \textit{n} is an input to the main program of each C++ source file.

\subsubsection{Transfer Schemes}
\label{simple_transfer_scheme}

For Linear scenarios, we have three options to transfer the data structures to the device: 

\begin{enumerate}
\item \textit{UVM:} Targeting NVidia GPUs, we utilized UVM for memory allocations. UVM allows developers to allocate memories that are accessible by both host and device. The PGI compiler provides UVM allocations with \texttt{-ta=tesla:managed} flag at the compile time for every memory allocation requests (\texttt{malloc}s) by the application.
\item \textit{Marshalling data structures:} We developed a method to enable the marshalling/demarshalling of structures at the run time of the application using \texttt{acc\_attach}/ \texttt{acc\_detach} API methods in OpenACC. Algorithm~\ref{marshalling_algo} shows the steps our implementation takes to implement the marshalling. At the beginning, developers determine how big the whole tree is (the main data structure with all of its nested data structures). Then, we allocate as much memory. Afterwards, any subsequent memory allocation requests from the program are responded by returning next available space from our allocated buffer. These steps compacts all the allocated memories into a contiguous space in the memory. This approach is the ideal case for transferring a complicated data structure tree in one batch instead of multiple batches per every structure. After transferring the whole buffer to the device, we have to call \texttt{acc\_attach} on each pointer on the device so that the pointers on the device point to a correct memory address. The demarshalling process is performed exactly in the reverse order of the marshalling algorithm. It is highly probable that the implementations of deep copy in different compilers follow similar marshalling approach. 
\item \textit{\texttt{pointerchain}:} Finally, we will investigate the effectiveness of our proposed directive as described in Section~\ref{pointerchain_directive_sec}. 
\end{enumerate}

\input{include/marshalling_algo.tex}

\subsubsection{Layout Schemes}

Three separate layout schemes are introduced for our Linear scenario. 
The layout schemes differ in whether the \textit{A}~arrays in Figure~\ref{fig:simple_approach_diagram} are allocated or not, and  whether they will be transferred to the device and utilized or not. 

\begin{enumerate}
\item \textit{allinit-allused}: In this scheme, all the \textit{A}~arrays in all levels allocate \textit{n} elements and they are accessed on the GPU. Our kernel scales all elements of the \textit{A}~arrays with an arbitrary number. This layout scheme helps us understand the efficiency of each transfer scheme when a full deep copy is inevitable. 
\item \textit{allinit-LLused}: Similarly, we allocate \textit{n} elements for all the \textit{A}~arrays, however only the \textit{A}~arrays of the last level is utilized within a kernel on the device. This scheme helps us understand how selective deep copy improves the performance when the kernels target only a subset of data structures on the device.
\item \textit{LLinit-LLused}: In this scheme, only the \textit{A}~array in the last-level~($L_k$) allocates memory space. This scheme helps us understand which transfer scheme performs the best in a long chain of pointers. This is a dominant scheme in scientific applications like molecular dynamics simulations~\cite{ghane2019pointerchain}. 
\end{enumerate}

\subsubsection{Data Size}
The amount of data generated by our tree of data structures for each layout scheme, as shown in Figure~\ref{fig:simple_approach_diagram}, is as following. For the \textit{allinit-allused} and \textit{allinit-LLused} cases, the size of our configuration, as a function of \textit{n} and \textit{k}, is:

\vspace{-1em}

\begin{equation}
\label{eq_simple_scenario}
\begin{split}
DataSize(k,n) & = \sum_{i=1}^{k} (24 + 8 n) \\
& = 24 k + 8 n k
\end{split}
\end{equation}

\noindent
where $24$ is the size of the $L_i$ structures and $8$ is the size of an element in $A$ in bytes (for double-precision floating-point numbers).

For \textit{LLinit-LLused} case, the data size can be computed as following:

\vspace{-1em}

\begin{equation}
\begin{split}
DataSize(k,n) & = \sum_{i=1}^{k} 24 + 8 n \\
& = 24 k + 8 n
\end{split}
\end{equation}

\subsection{Dense Scenario}

\begin{figure}[h]
\centering
\includegraphics[width=1\columnwidth]{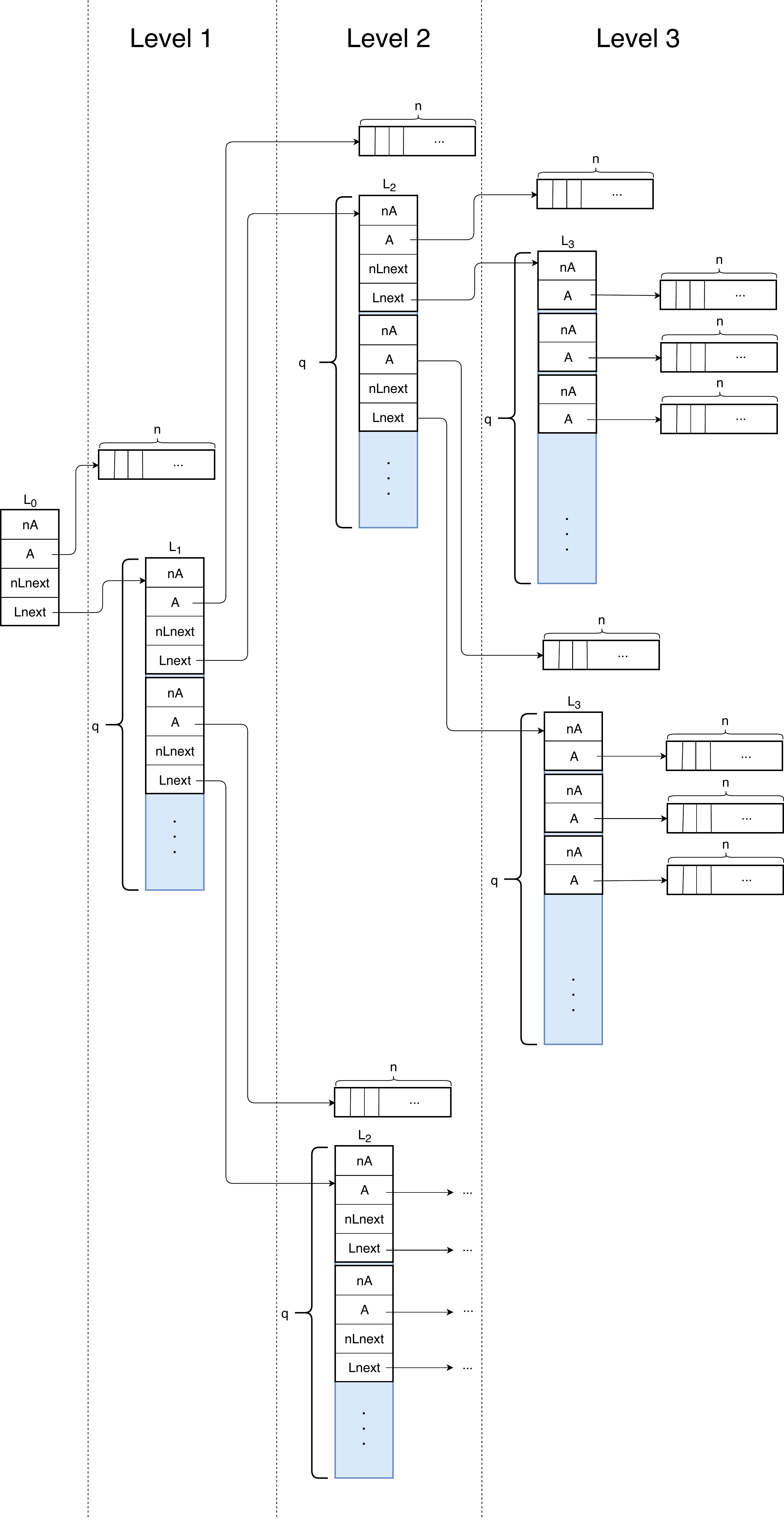}
\caption{The Dense case scenario. The three dots show recursive nature of the data structure.}
\label{fig:dense_diagram}
\vspace{-1em}
\end{figure}

In the dense scenario, the intermediate pointers are an array of objects instead of a single object. 
Figure~\ref{fig:dense_diagram} illustrates the dense scenario. This configuration provides a dense tree of data, which the size of the data will grow exponentially with small changes in both parameters in our design. The parameter~\textit{q} describes number of elements in the intermediate arrays~$L_i$, and the parameter~\textit{n} determines the number of elements in the \textit{A}~arrays.

\subsubsection{Transfer Scheme}
In comparison to the Linear scenario, transferring the data structure tree represented in Figure~\ref{fig:dense_diagram} is more complicated. For marshalling and \texttt{pointerchain} approaches, an extra work is required to make the intermediate pointers legal on the device so that they could be derefernced correctly. In cases similar to Dense, utilizing the \texttt{pointerchain} directive to perform a \textit{full deep copy} operation is not a viable option due to the increasing number of intermediate pointers, which grows exponentially in this case. 

We utilize UVM, marshalling, and \texttt{pointerchain} to transfer the data structure tree to the device similar to the Linear scenario. Each scheme is described in details in Section~\ref{simple_transfer_scheme}.

\subsubsection{Layout Scheme}

In the Dense scenario, we will choose an arbitrary index of each intermediate array $L_i$ (in our case, the last element of the array) and transfer the associated \textit{A}~array to the device to perform our computational kernel. For instance, for the configuration shown in Figure~\ref{fig:dense_diagram}, the kernel that we target to parallelize will look like Listing~\ref{listing:twolines}, where \texttt{q} is the number of elements in the intermediate arrays~$L_i$, and \texttt{a0} is the main structure at the first level.

\vspace{0.5em}
\input{include/two_lines.tex}
\vspace{0.5em}

\subsubsection{Data Size}

The amount of data generated by the data structure tree in the Dense scenario, as shown in Figure~\ref{fig:dense_diagram}, is very sensitive to the input parameters, $q$ and $n$. Small changes in these parameters leads to significant increases in the data size. Equation~\ref{dense_equation} shows the amount of data generated in bytes for our configuration in recursive form:

\vspace{-.5em}

\begin{equation}
\label{dense_equation}
\begin{split}
DataSize(q, n, D) & = 24 + 8 n + \\
                    & q \times DataSize(q, n, D-1) \\
DataSize(q, n, 0) & = 12 + 8 n
\end{split}
\end{equation}

\noindent
where 24 is the size of $L_i$ structures, 8 is the size of each element in array~\textit{A}, $q$ is the length of the intermediate arrays, and $D$ is the depth of our nested data structure. $DataSize(q, n,\\ 0)$ refers to the size of our last-level data structures (the L3 structures in Figure~\ref{fig:dense_diagram}). For our experiments in this paper, we set the maximum value of $D$ to $3$. Please note that the last-level data structure is half of the original structure in size.

%% file: include/marshalling_algo.tex

\begin{algorithm}[t]
  \small
  \caption{Marshalling algorithm}
  \label{marshalling_algo}
  \begin{algorithmic}[1]
	\Function{marshallize}{struct}
	
      \State $\algvar{n} \assign \Call{determineTotalBytes}{struct}$ 
	  
      \State $\algvar{buff} \assign $ Allocate n bytes buffer on heap 

      \State $\algvar{requestList} \assign [] $
      
      \For{memory allocation of size $\algvar{w}$}
        \State Append the allocation request to the $\algvar{requestList}$
        \State Return a pointer to $\algvar{w}$ bytes from $\algvar{buff}$
      \EndFor

	  \State Transfer $\algvar{buff}$ to the device
	  
      \For{req in $\algvar{requestList}$}
        \State $\Call{acc\_attach}{req}$
      \EndFor
      
    \EndFunction
  \end{algorithmic}
\end{algorithm}

%% file: include/two_lines.tex


\lstset{
language=C,
basicstyle=\small\sffamily,
numbers=left,
numberstyle=\small,
frame=l,
columns=fullflexible,
showstringspaces=false,
tabsize=2,
morekeywords={
    omp,
    parallel,
    acc,
    gecko,
    loctype,
    location,
    memory,
	\#pragma    
  },
  otherkeywords={\#pragma},
}

\lstset{
  emph={hierarchy},
  emphstyle={\color{black}\bfseries\itshape}
}

\begin{lstlisting}[caption={The scaling kernel used in our Dense scenario, where $q$ is the number of elements in the intermediate arrays~$L_i$.},label=listing:twolines]
for(int i=0;i<N;i++) 
    a0->Lnext[q-1].Lnext[q-1].Lnext[q-1].A[i] *= scale;
\end{lstlisting}

%% file: 05-evaluation.tex
\section{Experimental Setup}
\label{evaluation}

We performed our experiments on a diverse range of hardware and collected the results. Located at the University of Houston, Sabine~\cite{sabine} clusters host HPE compute nodes. 
Each systems are equipped with two Intel Xeon E5-2680v4 CPUs, with 28 logical cores, and 256GB host RAM. 
Sabine has both NVidia P100 and V100 GPU architectures. The P100 systems have 16GB global memory with 4MB L2 caches. The V100 GPUs also have 16GB global memory while their L2 caches are 6MB. Our software environment, for both system, include the PGI compiler ~18.4.




For the Linear scenario, we developed a Python script that accepts an integer number, $count$, as input and generates a set of source codes in C++ for $k \in [2, count]$. Each source code is a stand-alone application. The data structure tree depicted in Figure~\ref{fig:simple_approach_diagram} is generated statically for each $k$ to allow the compilers apply optimizations on the source codes efficiently. For each $k$, our script generates nine files: three transfer schemes by three layout schemes. As an example, suppose we pass 10 to our Python script. Then, total files generated by our script is 81 ($(count - 2 + 1) \times 3 \times 3=81$).

For the Dense scenario, we developed three different transfer schemes~(UVM, marshalling, and \texttt{pointerchain}) to perform the selective deep copy. Each scheme accepts two inputs, $n$ and $q$, which they were previously described in Section~\ref{model}.

Algorithm~\ref{main_program_algo} displays the steps that each benchmark application takes. At the beginning of the application, we allocate the memory for our data structure tree. We, then, initialize them with arbitrary values. Then, we will transfer the whole data structure to the device based on the various transfer schemes explained in details in Section~\ref{model}. We will run a kernel on our tree. The kernel scales every elements of the array~\textit{A} by a constant value. Based on the chosen layout scheme, whether it is \textit{allused} or \textit{LLused}, all or last-level \textit{A}-arrays are scaled, respectively. After running the kernel, we will transfer the tree back to the host and check the results.

\input{include/main_program_algo.tex}

For both Linear and Dense scenarios, we will measure two different metrics: (a)~the wall-clock time of the whole application, (b)~the kernel execution time. The wall-clock time is measured to investigate the effect of each transfer scheme on each different scenario. 
The kernel execution time is measured to give us an insight about how different data layouts affect kernel's performance. Not only the execution time, but also total instructions generated by the compiler will be affected with different transfer schemes.


We used Google~Benchmark~\cite{googlebenchmark} to measure the execution time (i.e., the kernel and the wall-clock time). It is a lightweight, powerful framework to benchmark functions. Through a set of preliminary testing, the framework learns how many iterations is required to be performed so that we get a consistent result within a low error margin at the end. Each test case is implemented as a function, and then, the whole function is benchmarked with Google Benchmark. For the results of the kernel time, we benchmarked only the kernel computations on Step~4 (line~5) of Algorithm~\ref{main_program_algo}.

%% file: include/main_program_algo.tex
\begin{algorithm}[t]
  \small
  \caption{Main program steps}
  \label{main_program_algo}
  \begin{algorithmic}[1]
	\Function{main}{argc, argv}
	
	    \State 1- Allocate memory for whole tree structure

	    \State 2- Initialize the tree

	    \State 3- Transfer the tree to the device with a transfer scheme

	    \State 4- Run the kernel once

	    \State 5- Transfer the tree back to the host
	    
	    \State 6- Check the results
	    
	    \State 7- Measure the wall-clock time
	
    \EndFunction
  \end{algorithmic}
\end{algorithm}

%% file: 06-result.tex
\section{Results}
\label{results}

We performed our experiments that were designed in Section~\ref{model} on the Sabine systems~(P100 and V100). Results are provided in this section.

\subsection{Linear Scenario}

\input{include/simple_total_time_figure.tex}

\input{include/simple_compute_time_figure.tex}

\input{include/mem_size_simple_scenario.tex}

We measured the wall-clock and kernel time of the experiments designed for the Linear scenario. Figure~\ref{simple_scheme_total_time} shows the wall-clock time for different number of levels and different layout schemes. Results are normalized with respect to the UVM approach.

\subsubsection{Wall-clock time}

Results for the \textit{allinit-allused} transfer scheme reveal how increasing parameter~$n$ leads to the performance loss for all values of $k$. As we increase the total size of the tree (increasing both $n$ and $k$), there is no performance loss when UVM is utilized, and it has a chance to be a viable option in comparison to other methods. 
Furthermore, UVM is a feasible approach to transfer data between host and device when applications are dealing with huge amount of data.
It provides developers more productivity with the same level of performance when we are targeting huge data. 
However, when $n$ is not moderately huge (for $n < 10^5$) and the chain length ($k$) is small, marshalling and \texttt{pointerchain} outperform UVM. 
Furthermore, there is no subtle difference between different architectures (P100 and V100) for the \textit{allinit-allused} scheme.

On the other hand, the \textit{allinit-LLused} scheme is more susceptible to the transfer scheme rather than the underlying architecture. As $n$ increases in the size, the gap between marshalling and \texttt{pointerchain} increases. For larger $k$ values, \texttt{pointerchain} outperforms the marshalling and UVM. Thus, \texttt{pointerchain} is the better option for a deep copy operation in comparison to the other two options when we are dealing with huge data sets.
As $k$ increases, the marshalling scheme performs worse while the performance of \texttt{pointerchain} is not affected and remains constant. 
There is no notable difference between different architectures, and the transfer schemes determines the performance. It is the underlying data transfer medium, in our case the PCI-E bus, that determines the upper bound of the performance.

Finally, for the \textit{LLinit-LLused} scheme, UVM has the worst performance results. The results show how in cases that our kernel targets an array at the last-level data structure, utilizing either marshalling or \texttt{pointerchain} leads to better performance results. The \texttt{pointerchain} scheme shows promising results when $n < 10^5$. However, for $n > 10^5$, the architecture design determines the winner. The V100 architecture shows 2X improvements in performance for marshalling and \texttt{pointerchain} schemes, however, P100 was able to show 1.25X improvement. For all values of $k$ and $n$, \texttt{pointerchain} performed better than marshalling.

\subsubsection{Kernel execution time}

Figure~\ref{simple_scheme_compute_time} shows the normalized kernel time with respect to UVM for different level count and different layout schemes. There is no subtle difference among different transfer schemes, different layout schemes, and different architectures. Mostly, for all values of $n$ and $k$, all results follow the same trend. 
However, we observe the best performance when $n \in [10^4,10^6]$.

Table~\ref{data_size_table_for_simple_scenario} shows the total size of our data structure tree as we change $k$ and $n$. For all $k$s, while $n < 10^5$ the whole data fits in the L2 cache of P100 and V100 GPUs. As we increase $n$, the L2 cache is not big enough anymore, which results in the mandatory cache eviction process, subsequently, we lose performance. This is the reason that we observe an increasing trend in the execution time in Figure~\ref{simple_scheme_compute_time}.
This confirms our finding: \textit{when we are dealing with the data structures with huge sizes, there is no subtle difference in performance between UVM and other transfer schemes for complex data structures.}

\input{include/mem_size_dense_scenario.tex}

\input{include/dense_total_compute_figure.tex}

\subsection{Dense Scenario}

We measured the wall-clock and kernel time of the experiments designed for the Dense scenario. Figure~\ref{dense_scheme_total_compute} shows the normalized wall-clock time and kernel time with respect to UVM for different level count and different layout schemes.

\subsubsection{Wall-clock time}

The key factor that determines the performance of the whole application is the transfer scheme. The \texttt{pointerchain} scheme performs consistently better in comparison to the marshalling. In cases like $n = 10$ and $n = 100$, \texttt{pointerchain} basically shows two orders of magnitude performance improvements in comparison to marshalling. In such cases, UVM shows close to 10X improvement over marshalling.

However, as $q$ increases, the performance gap between \texttt{pointerchain} and marshalling shrinks. Moreover, Figure~\ref{dense_scheme_total_compute} shows how in the Dense scenarios, the underlying architecture does not have any contributions to the performance. It is the transfer scheme that determines the performance. The reason behind such performance deficiency of the marshalling scheme is the extra job required to be done to ensure the pointer consistency on the device. For each pointer, we are required to fix the address in the structure to point to a correct location on the memory space of the device.

\input{include/ptx-count-table.tex}

\subsubsection{Kernel execution time}

Figure~\ref{dense_scheme_total_compute} also demonstrates the performance of the kernel with respect to different transfer schemes introduced in Section~\ref{model}. Despite no subtle differences, the marshalling scheme leads to more performance friendly data layout in comparison to \texttt{pointerchain} on both architectures. While kernels that are executed on the marshalled data perform better than their UVM counterparts, the \texttt{pointerchain} scheme suffers some performance loss. Consequently, in cases that a kernel is executed multiple times on the same data, the data layout of the marshalling scheme results in a better performance. Such an effect is due to the cache friendly layout of our implementation for marshalling. The marshalling scheme places the arrays as close as possible to the pointers that points to them, however, this is not necessarily the case for the \texttt{pointerchain} scheme. In \texttt{pointerchain}, the arrays are scattered around the global memory of GPUs and they do not necessarily reside in the same memory page as the pointer itself. 

\subsection{Instruction Count}

The process of dereferencing pointers generates a set of instruction to retrieve the effective address of the pointer. For Tesla V100, the PGI compiler generates 2 instructions per each dereference operation: 
1) an instruction to load the address from global memory to a register (\texttt{\justify ld.global.nc .u64}); 
2) an instruction to convert the virtual address to a physical address on the device (\texttt{\justify cvta.to.global.u64}). 
For every chain, the processor has to execute above instructions to extract the effective address.

Table~\ref{instruction_count_simple} and \ref{instruction_count_dense} show total number of generated instructions by the PGI compiler for the Linear and Dense scenarios, respectively. To count number of instructions, we generated the PTX files by enabling the \texttt{keep} flag at compile time (\texttt{-ta=tesla:cc70,keep}). Then, we counted number of lines~(LOC) in the generated PTX file.

The results for the Linear scenario, as shown in Table~\ref{instruction_count_simple}, reveals up to 31\% reduction in the generated code for GPUs. 
The LOC for the \texttt{LLused} schemes remains constant since we are basically reducing any pointer chains in our application to one pointer. However, for UVM and marshalling schemes, as $k$ increases, total generated code for them also increases as well since we have to dereference the chain of pointers. For the \textit{allinit-\textbf{LLused} and LLinit-\textbf{LLused}} schemes, one can observe how the LOC increases by two lines between two consecutive $k$s. For the \textit{allinit-allused} scheme, since we are dealing with multiple pointer chains, the trend is not linear, however we save more instructions in this case. Table~\ref{instruction_count_dense} shows similar results for the Dense scenario. We have two observations: 1)~The marshalling scheme did not increase number of instructions with respect to UVM. 2)~\textbf{\texttt{pointerchain}} led to 25\% reduction in generated instructions.

%% file: include/simple_total_time_figure.tex


\begin{figure*}[t]
    \centering
    \subcaptionbox{\textit{allinit-allused} scheme}[.32\linewidth][c]{%
        \includegraphics[width=.34\linewidth]{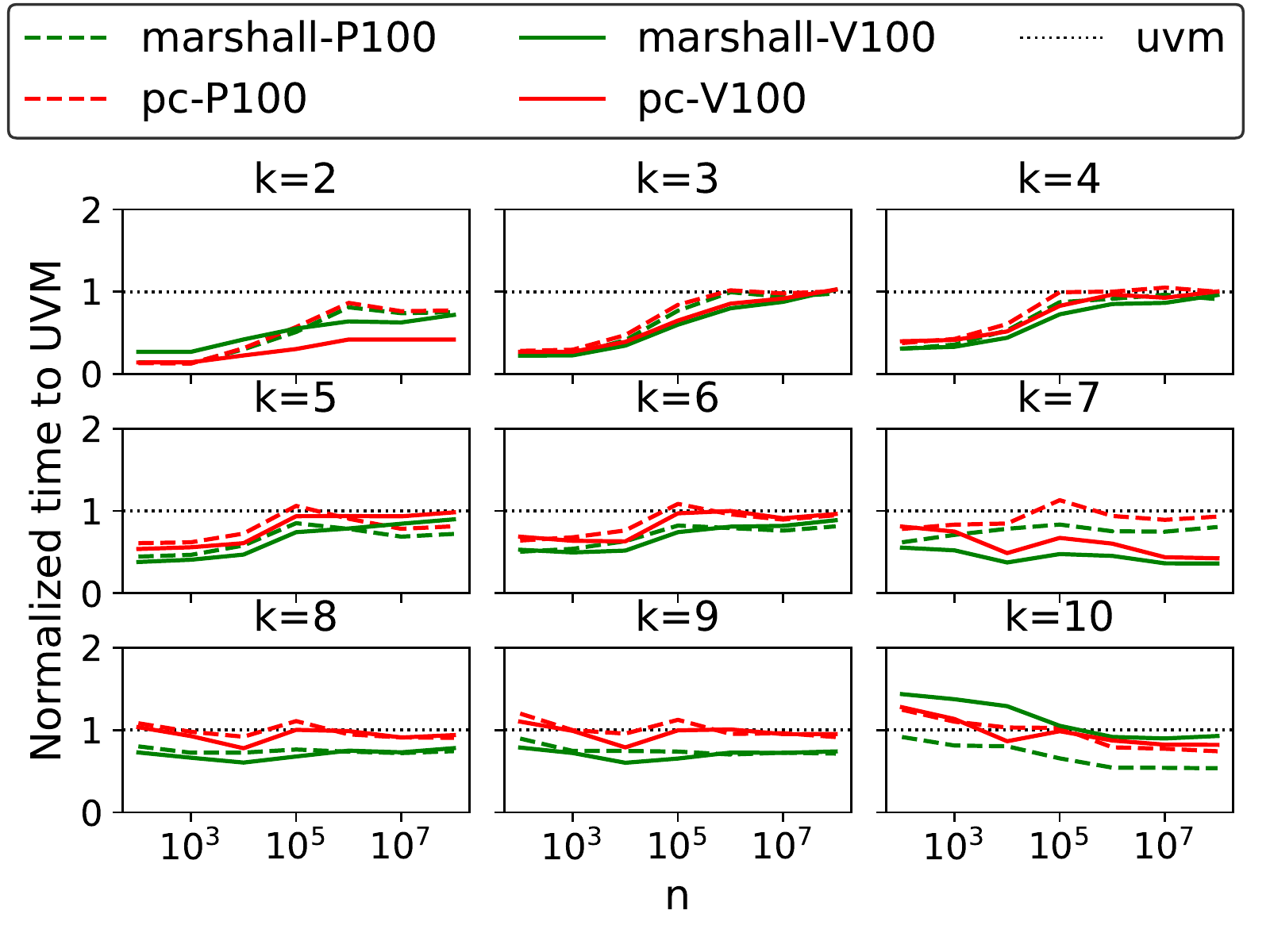}
    }\quad
    \subcaptionbox{\textit{allinit-LLused} scheme}[.32\linewidth][c]{%
        \includegraphics[width=.34\linewidth]{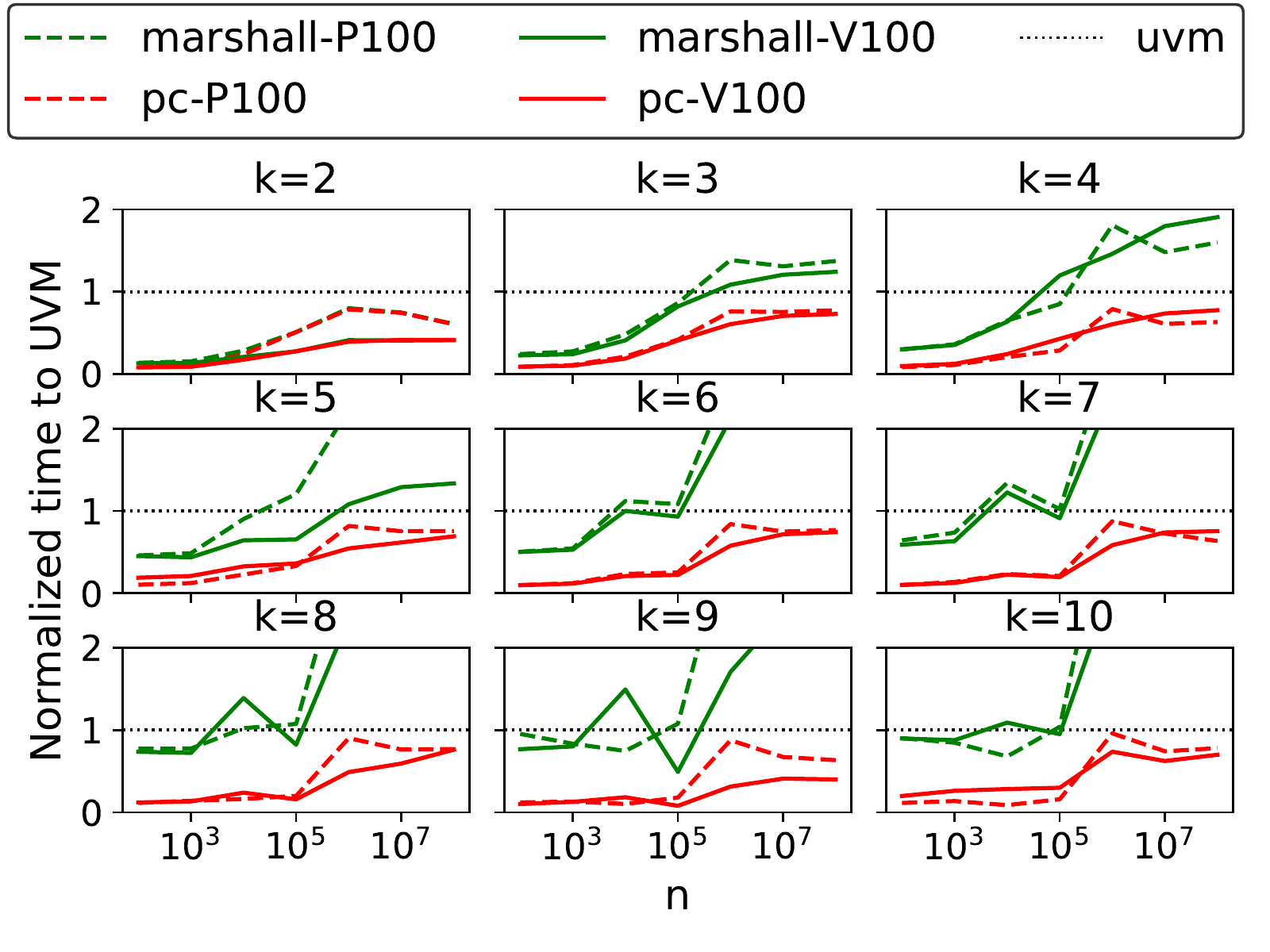}
    }\quad
    \subcaptionbox{\textit{LLinit-LLused} scheme}[.32\linewidth][c]{%
        \includegraphics[width=.34\linewidth]{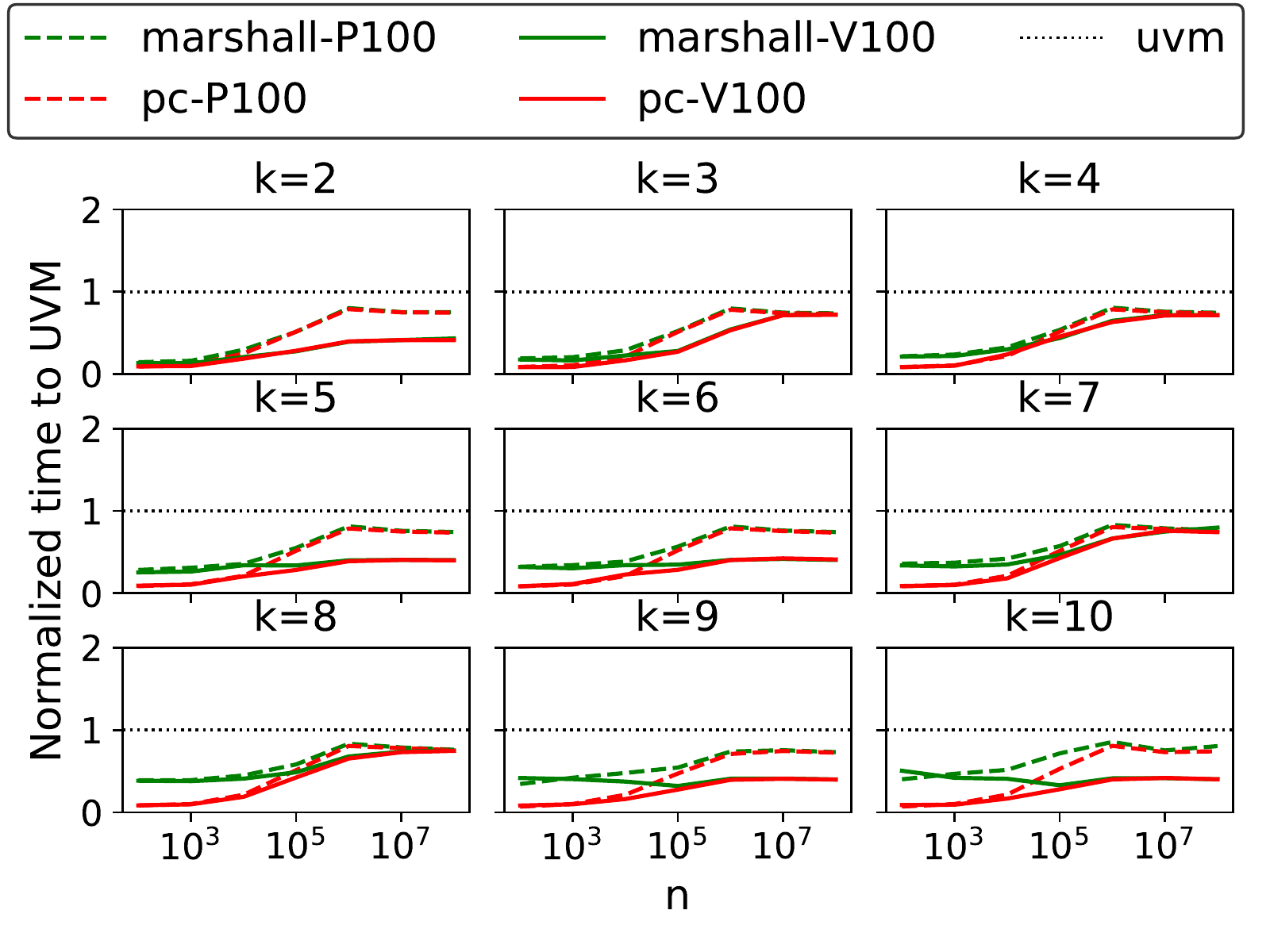}
    }


  \caption{Normalized wall-clock time with respect to UVM. Lower is better.}
  \label{simple_scheme_total_time}
\end{figure*}

%% file: include/simple_compute_time_figure.tex


\begin{figure*}[t]
    \centering
    \subcaptionbox{\textit{allinit-allused} scheme}[.32\linewidth][c]{%
        \includegraphics[width=.34\linewidth]{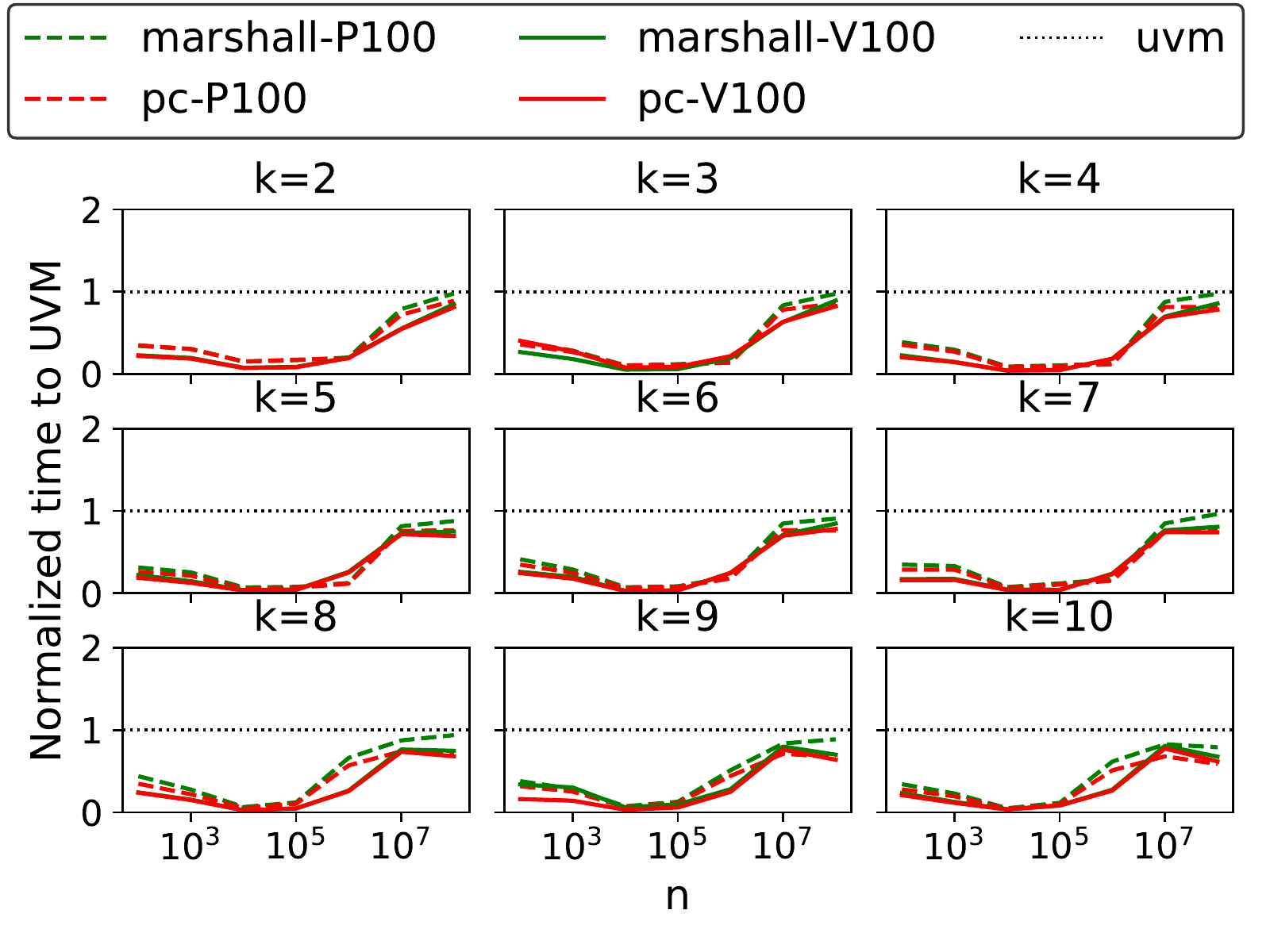}
    }\quad
    \subcaptionbox{\textit{allinit-LLused} scheme}[.32\linewidth][c]{%
        \includegraphics[width=.34\linewidth]{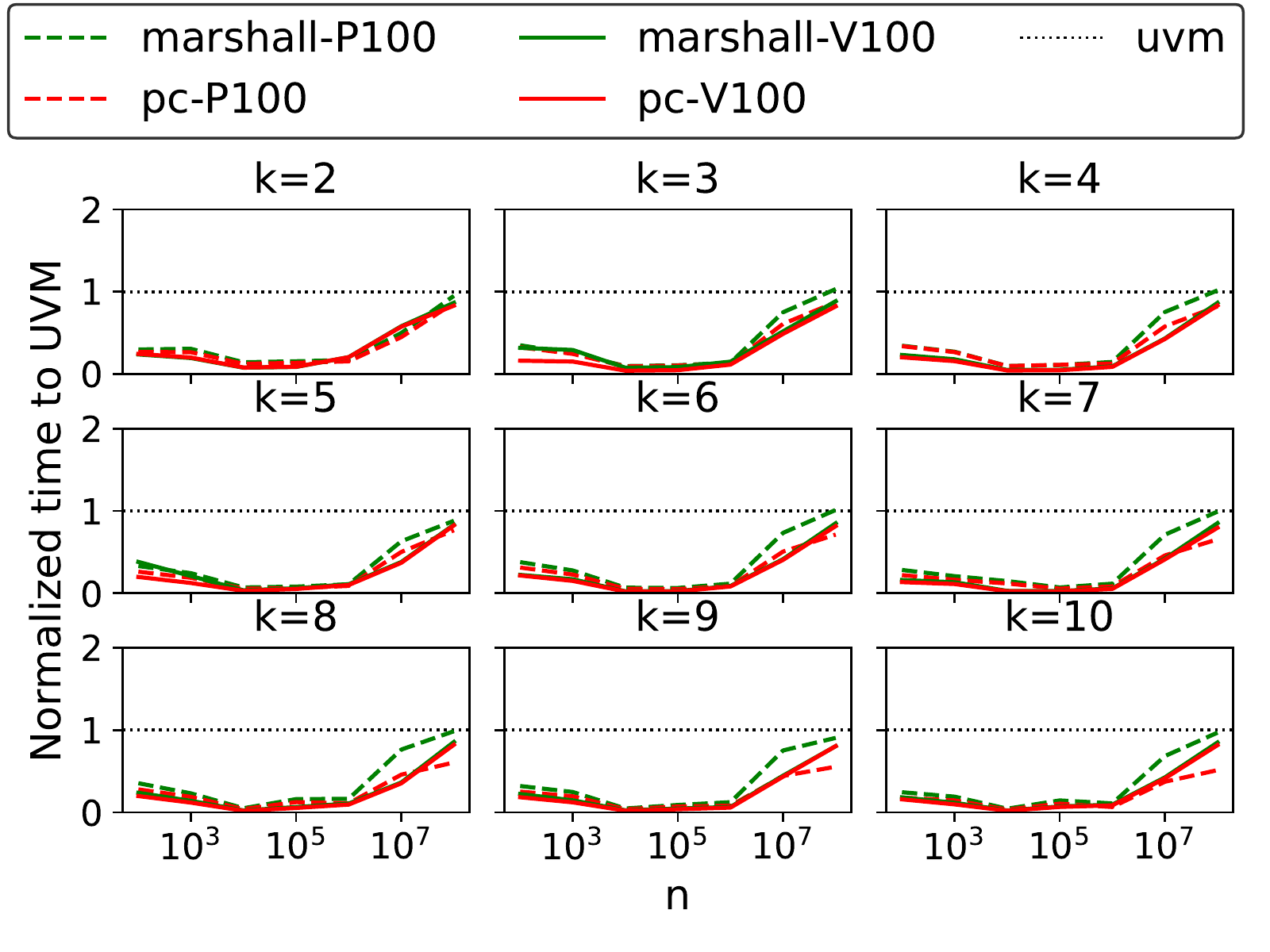}
    }\quad
    \subcaptionbox{\textit{LLinit-LLused} scheme}[.32\linewidth][c]{%
        \includegraphics[width=.34\linewidth]{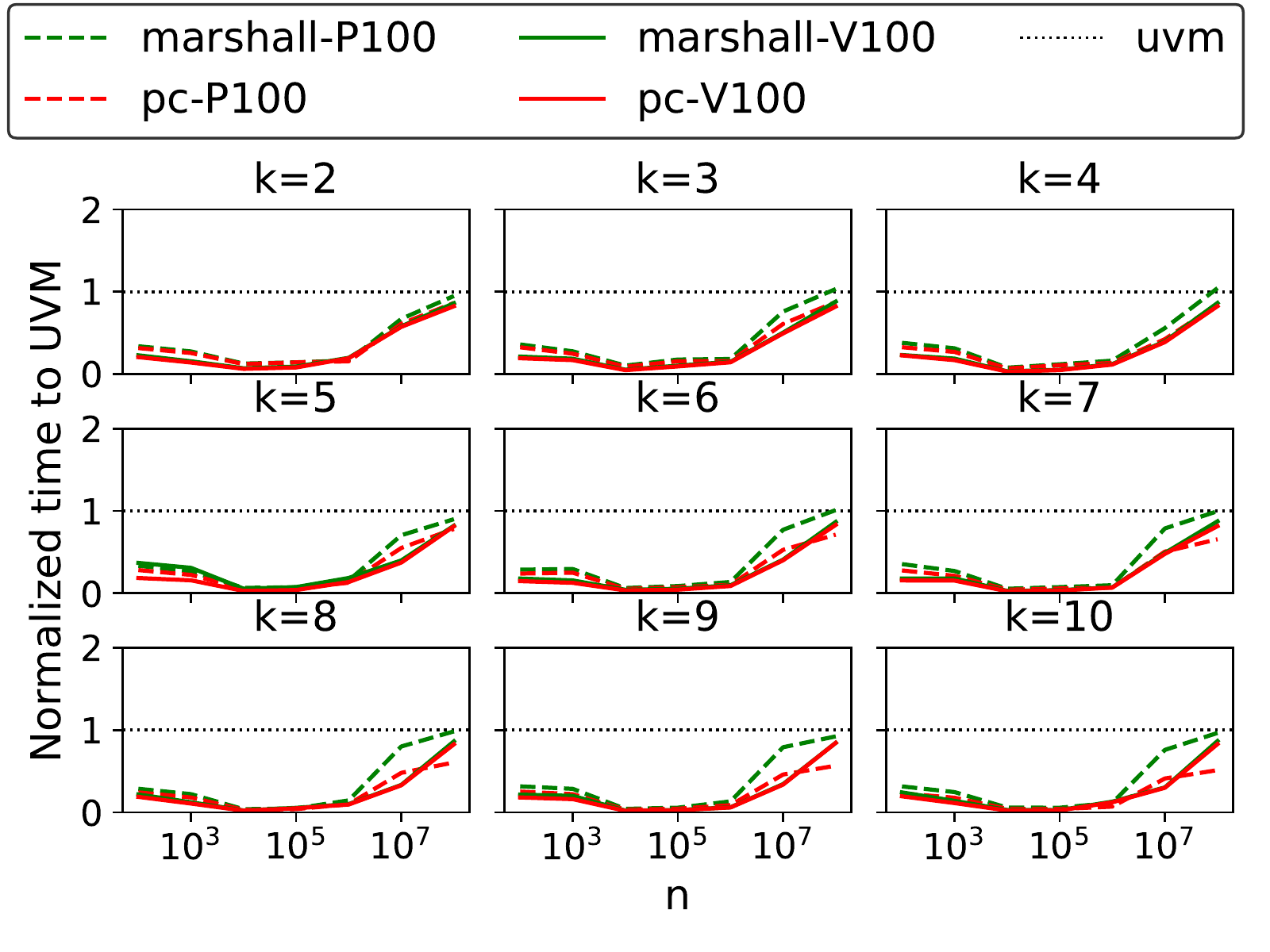}
    }


  \caption{Normalized kernel time with respect to UVM. Lower is better.}
  \label{simple_scheme_compute_time}
\end{figure*}

%% file: include/mem_size_simple_scenario.tex
\begin{table*}[t]
\caption{Total data size of our data structure tree as defined in the Linear scenario for the \textit{allinit-allused} scheme. We used Equation~\ref{eq_simple_scenario} to calculate these numbers. One can see how the data size increases as we increase $n$ and $k$. \textit{The first row is in KiloBytes, while the rest of the numbers are in MegaBytes.}}
\label{data_size_table_for_simple_scenario}
\vspace{-1em}
\footnotesize
\def\arraystretch{0.8}
\begin{tabular}{c|rrrrrrrrrr}
\toprule
          & \multicolumn{9}{|c}{k}             \\ \midrule
n         & 2 & 3 & 4 & 5 & 6 & 7 & 8 & 9 & 10 \\ \midrule
 $10^2$  & \textit{1.61 KB}  & \textit{2.41 KB}  & \textit{3.22 KB}  & \textit{4.02 KB}  & \textit{4.83 KB}  & \textit{5.63 KB}  & \textit{6.44 KB}  & \textit{7.24 KB}  & \textit{8.05 KB}  \\
  $10^3$  & 0.02 MB  & 0.02 MB  & 0.03 MB  & 0.04 MB  & 0.05 MB  & 0.05 MB  & 0.06 MB  & 0.07 MB  & 0.08 MB  \\
  $10^4$  & 0.15 MB  & 0.23 MB  & 0.31 MB  & 0.38 MB  & 0.46 MB  & 0.53 MB  & 0.61 MB  & 0.69 MB  & 0.76 MB  \\
  $10^5$  & 1.53 MB  & 2.29 MB  & 3.05 MB  & 3.81 MB  & 4.58 MB  & 5.34 MB  & 6.10 MB  & 6.87 MB  & 7.63 MB  \\
  $10^6$  & 15.26 MB  & 22.89 MB  & 30.52 MB  & 38.15 MB  & 45.78 MB  & 53.41 MB  & 61.04 MB  & 68.66 MB  & 76.29 MB  \\
  $10^7$  & 152.59 MB  & 228.88 MB  & 305.18 MB  & 381.47 MB  & 457.76 MB  & 534.06 MB  & 610.35 MB  & 686.65 MB  & 762.94 MB  \\
  $10^8$  & 1525.88 MB  & 2288.82 MB  & 3051.76 MB  & 3814.70 MB  & 4577.64 MB  & 5340.58 MB  & 6103.52 MB  & 6866.46 MB  & 7629.39 MB \\ \bottomrule

\end{tabular}
\end{table*}

%% file: include/mem_size_dense_scenario.tex
\begin{table*}[t]
\caption{Total data size of our data structure tree as defined in the Dense scenario. We used Equation~\ref{dense_equation} to calculate these numbers. }
\label{data_size_table_for_dense_scenario}
\vspace{-1em}
\footnotesize
\def\arraystretch{0.8}
\begin{tabular}{c|rrrrrrrrrr}
\toprule
          & \multicolumn{8}{c}{q}             \\ \midrule
n         & 2 & 4 & 6 & 8 & 10 & 12 & 14 & 16 \\ \midrule
 $10^1$  & 1.43 KB  & 7.88 KB  & 0.02 MB  & 0.05 MB  & 0.10 MB  & 0.17 MB  & 0.26 MB  & 0.39 MB  \\
  $10^2$  & 0.01 MB  & 0.07 MB  & 0.20 MB  & 0.45 MB  & 0.86 MB  & 1.46 MB  & 2.29 MB  & 3.39 MB  \\
  $10^3$  & 0.11 MB  & 0.65 MB  & 1.98 MB  & 4.47 MB  & 8.49 MB  & 0.01 GB  & 0.02 GB  & 0.03 GB  \\
  $10^4$  & 1.14 MB  & 6.49 MB  & 0.02 GB  & 0.04 GB  & 0.08 GB  & 0.14 GB  & 0.22 GB  & 0.33 GB  \\
  $10^5$  & 0.01 GB  & 0.06 GB  & 0.19 GB  & 0.44 GB  & 0.83 GB  & 1.40 GB  & 2.20 GB  & 3.26 GB \\ \bottomrule
  
\end{tabular}
\end{table*}

%% file: include/dense_total_compute_figure.tex


\begin{figure*}[t]
    \centering
    \subcaptionbox{\textit{Wall-clock time}}[.45\linewidth][c]{%
        \includegraphics[width=.45\linewidth]{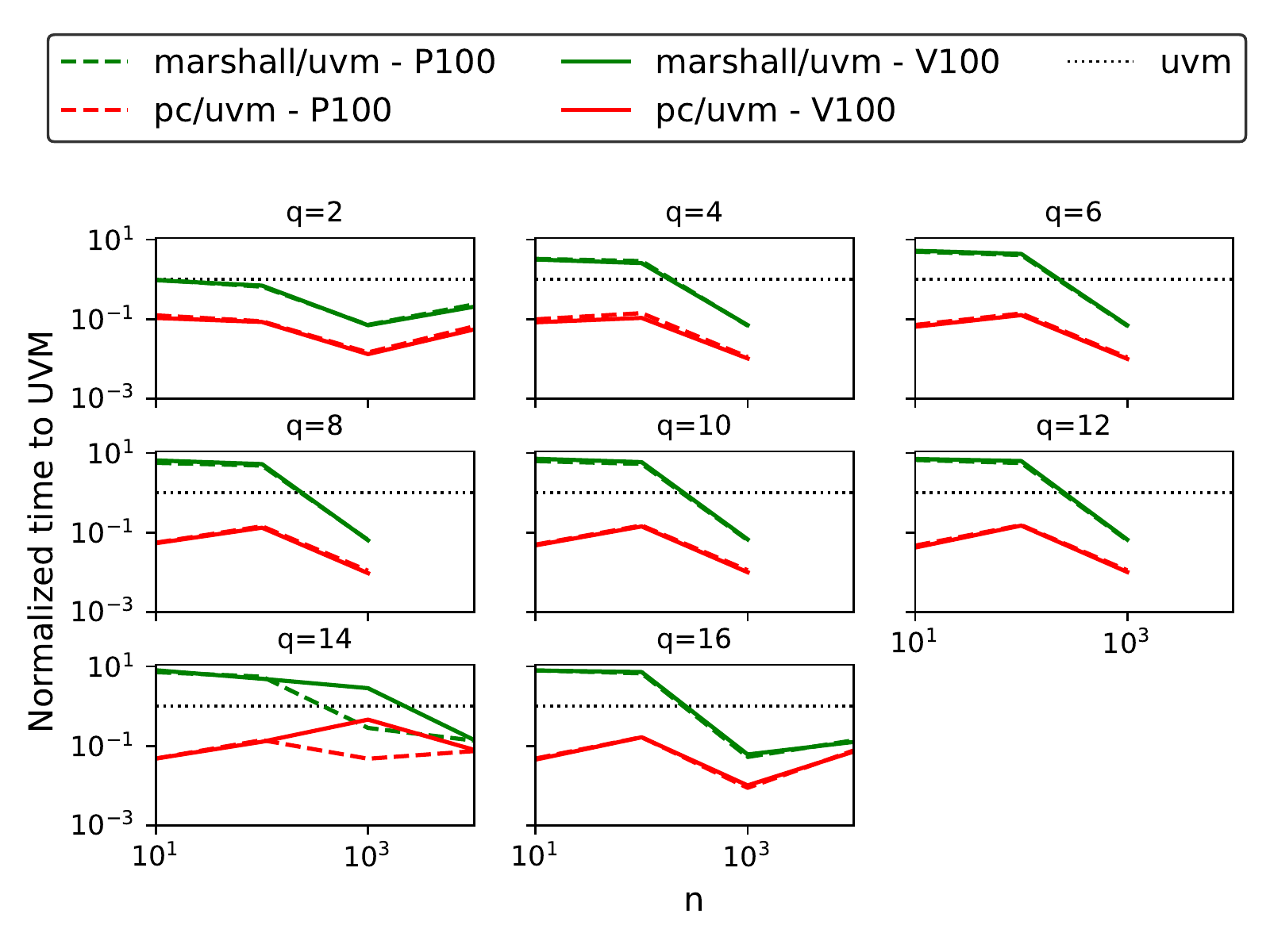}
    }\quad
    \subcaptionbox{\textit{Kernel time}}[.45\linewidth][c]{%
        \includegraphics[width=.45\linewidth]{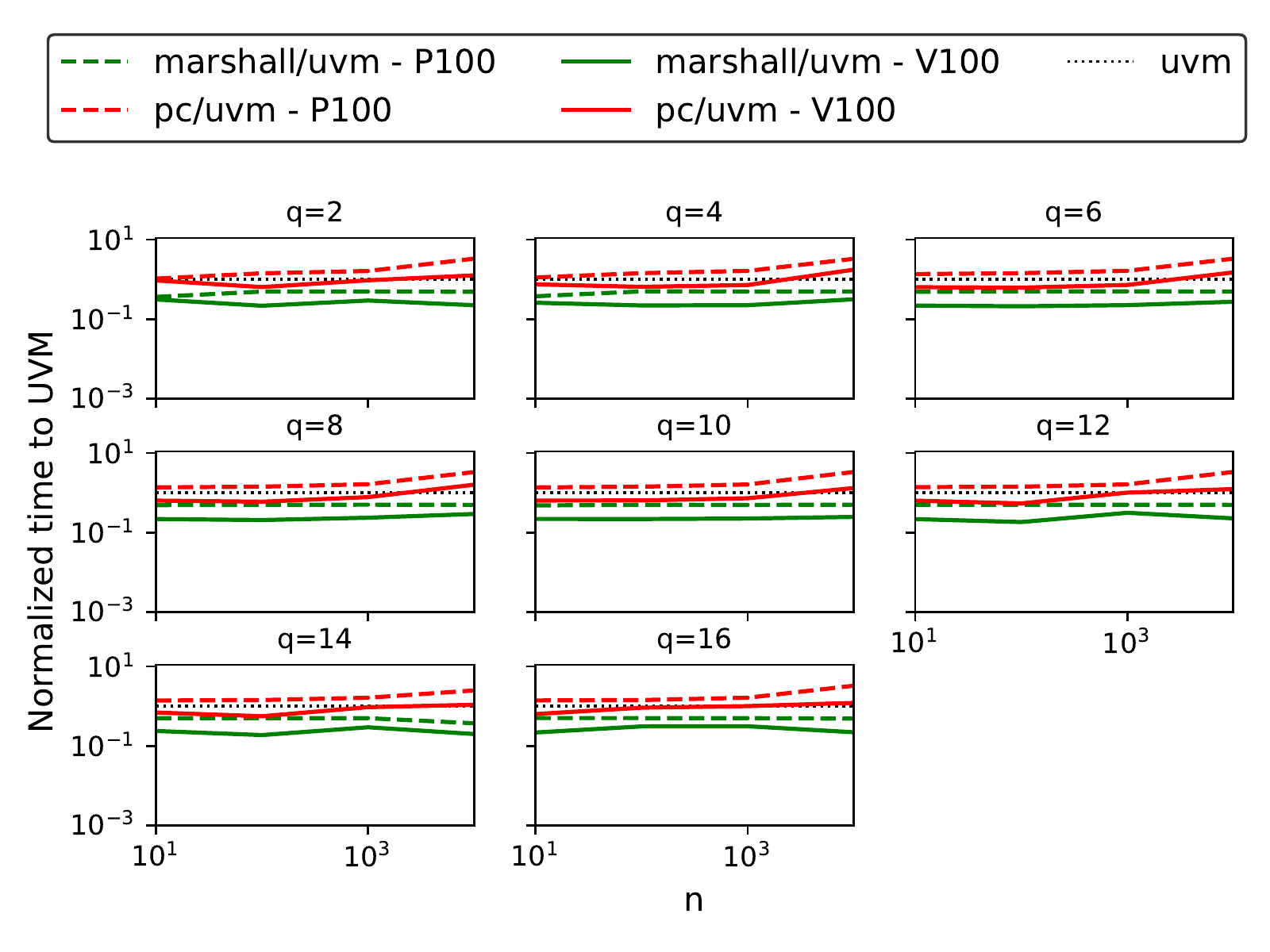}
    }\quad

  \caption{Normalized wall-clock time and kernel time to UVM for Dense scenario. The Y axes are in logarithmic scale. Lower is better.}
  \label{dense_scheme_total_compute}
  
\end{figure*}

%% file: include/ptx-count-table.tex
\begin{table*}[!htb]
\footnotesize
    \begin{minipage}[t]{.65\textwidth}
      \centering
                    \caption{Total instruction generated by the PGI compiler (for Tesla V100) for the Linear scenario. \textit{Mar.} and \textit{PC} refer to the marshalling and \textbf{\texttt{pointerchain}} schemes, respectively. The numbers in parentheses show the increase with respect to UVM.}
                    \label{instruction_count_simple}
                    \vspace{-1em}
                    \begin{tabular}{|c||c|c|c||c|c|c||c|c|c|}
                    \hline
                       & \multicolumn{3}{c||}{allinit-allused} & \multicolumn{3}{c||}{allinit-LLused} & \multicolumn{3}{c|}{LLinit-LLused} \\ 
                       \hline
                       \hline
                    k  & UVM    & Mar. (\%)   & PC (\%)       & UVM    & Mar. (\%)   & PC (\%)      & UVM   & Mar. (\%)   & PC (\%)      \\ \hline
                    2  & 62     & 62 (0\%)    & 60 (-3\%)     & 62     & 62 (0\%)    & 60 (-3\%)    & 62    & 62 (0\%)    & 60 (-3\%)    \\ \hline
                    3  & 70     & 70 (0\%)    & 67 (-4\%)     & 64     & 64 (0\%)    & 60 (-6\%)    & 64    & 64 (0\%)    & 60 (-6\%)    \\ \hline
                    4  & 78     & 78 (0\%)    & 74 (-5\%)     & 66     & 66 (0\%)    & 60 (-9\%)    & 66    & 66 (0\%)    & 60 (-9\%)    \\ \hline
                    5  & 88     & 88 (0\%)    & 81 (-8\%)     & 68     & 68 (0\%)    & 60 (-12\%)   & 68    & 68 (0\%)    & 60 (-12\%)   \\ \hline
                    6  & 100    & 100 (0\%)   & 88 (-12\%)    & 70     & 70 (0\%)    & 60 (-14\%)   & 70    & 70 (0\%)    & 60 (-14\%)   \\ \hline
                    7  & 114    & 114 (0\%)   & 95 (-17\%)    & 72     & 72 (0\%)    & 60 (-17\%)   & 72    & 72 (0\%)    & 60 (-17\%)   \\ \hline
                    8  & 130    & 130 (0\%)   & 102 (-22\%)   & 74     & 74 (0\%)    & 60 (-19\%)   & 74    & 74 (0\%)    & 60 (-19\%)   \\ \hline
                    9  & 148    & 148 (0\%)   & 109 (-26\%)   & 76     & 76 (0\%)    & 60 (-21\%)   & 76    & 76 (0\%)    & 60 (-21\%)   \\ \hline
                    10 & 168    & 168 (0\%)   & 116 (-31\%)   & 78     & 78 (0\%)    & 60 (-23\%)   & 78    & 78 (0\%)    & 60 (-23\%)   \\ \hline
                    \end{tabular}
    \end{minipage}%
    \hfill
    \begin{minipage}[t]{.3\textwidth}
      \centering
                    \caption{Total instruction generated by the PGI compiler (for Tesla V100) for the Dense scenario. \textit{Mar.} and \textit{PC} refer to the marshalling and \textbf{\texttt{pointerchain}} schemes, respectively. The numbers in parentheses show the increase with respect to UVM.}
                    \label{instruction_count_dense}
                    \begin{tabular}{|c|c|c|c|}
                    \hline
                          & UVM & Mar. (\%)   & PC (\%)   \\ \hline
                    Dense & 80  & 80 (0\%) & 60 (-25\%) \\ \hline
                    \end{tabular}
    \end{minipage} 
\end{table*}

%% file: 07-relatedwork.tex
\section{Related work}
\label{relatedwork}

Modern HPC applications and simulation frameworks make extensive use of deeply nested data structures in their design and source code~\cite{lindahl2001gromacs,pearlman1995amber,plimpton1995fast,icon2016}. In such cases, patterns depicted in Figure~\ref{fig:pointerchain} happens frequently in their source code, and this requires extensive care to ensure the data consistency in a heterogeneous environment with different memory spaces. We need a deep copy of the data structures between different spaces in such environments. 

Deep copy has been a challenging task for the HPC developers for the past couple of years. 
Technical report~(TR-16-1)~\cite{deepcopyTR2016} was the first attempt to formulate and propose a solution to the deep copy problem based on a real HPC application~(ICON~\cite{icon2016}). Cray~\cite{beyer2014transferring} proposes the utilization of \texttt{policy} and \texttt{shape} within the definition of the data structures to support both \textit{selective} and \textit{full} deep copy. However, it was only supported by their compiler. The PGI compiler has recently started to support deep copy in their latest compiler as a part to support a draft implementation of OpenACC~3.0. However, at the time of writing this paper, we did not have access to their latest version of the PGI compiler. The proposed solution by above-mentioned vendors are not the same and they differ in their approach. 

NVidia, on the other hand, introduced its UVM technology~\cite{landaverde2014uvm} to eliminate the need for manually updating different memories of a heterogeneous system. The underlying CUDA library will track the dirty pages on the memory subsystems and provides the most up-to-date version of a memory page on the device requesting it. However, unlike deep copy approaches, UVM requests happens at arbitrary times during executing an application and causes slow down when running an kernel. 

%% file: 08-conclusion.tex
\section{Conclusion}
\label{conclusion}

In this paper, we designed and implemented a benchmark suite for deep copy operations in a heterogeneous platform. We introduced two set of scenarios for different setups of developing an application: Linear and Dense scenarios. In short, Linear helps investigating the effect of a sparse data structure tree, while Dense helps studying a very dense  data structure. Each scenario has a set of transfer and layout schemes. Transfer schemes determine how the whole tree of the data structure (the main structure with its nested ones) is transferred to and from the devices. The layout schemes determine how allocations are performed for each array within the data structures. 

In addition to the benchmark suite, we proposed \textbf{\texttt{pointerchain}} as a low-overhead, simple directive to address the \textit{selective} deep copy for nested data structures. Our results reveal how \texttt{pointerchain} outperforms current state-of-the-art approaches. In the Dense scenarios, the \texttt{pointerchain} performs orders of magnitude better than UVM by NVidia.